# N-V$_{Si}$-related center in non-irradiated 6H SiC nanostructure

Nikolay Bagraev[a], Eduard Danilovskii[a], Dmitrii Gets[a], Ekaterina Kalabukhova[b], Leonid Klyachkin[a], Anna Malyarenko[a], Dariya Savchenko[b,c] and Bella Shanina[b]

[a]*Ioffe Physical Technical Institute, Polytekhnicheskaya 26, 194021 St. Petersburg, Russia*
[b]*Lashkaryov Institute of semiconductor physics, NASU, Ukraine*
[c]*Institute of Physics, AS CR, Czech Republic*

**Abstract.** We present the first findings of the vacancy-related centers identified by the electron spin resonance (ESR) and electrically-detected (ED) ESR method in the non-irradiated 6H-SiC nanostructure. This planar 6H-SiC nanostructure represents the ultra-narrow p-type quantum well confined by the δ-barriers heavily doped with boron on the surface of the n-type 6H-SiC (0001) wafer. The EDESR method by measuring the only magnetoresistance of the 6H SiC nanostructure under the high frequency generation from the δ-barriers appears to allow the identification of the silicon vacancy centers as well as the triplet center with spin state S=1. The same triplet center that is characterized by the larger value of the zero-field splitting constant D and anisotropic g-factor is revealed by the ESR (X-band) method. The hyperfine (hf) lines in the ESR and EDESR spectra originating from the hf interaction with the $^{14}$N nucleus allow us to attribute this triplet center to the N-V$_{Si}$ defect.

## INTRODUCTION

Recently the N-V$_{Si}^-$ (S=1) defect in the diamond, 4H-and 6H-SiC structures became to be of interest as a qubit for quantum computing operations owing to the possibility to control the coherent spin processes at room temperature [1]. However, up to now the N-V$_{Si}$ defect and other silicon vacancy related centers were introduced into these wide gap semiconductors only by the neutron, proton and electron irradiation with subsequent thermal annealing. Therefore the preparation of samples containing the only N-V$_{Si}$ defects and isolated silicon vacancies for the realization of the qubit versions has been always difficult because of the different defects created under this irradiation which are able to give rise to non-radiative recombination.

In present work we demonstrate the results of the electron spin resonance (ESR) and electrically-detected (ED) ESR studies of the N-V$_{Si}$ defect and the isolated silicon vacancy centers which are formed in the course of the preparation of the planar 6H-SiC nanostructure without any previous or subsequent e- and n- irradiation.

## METHODS

The device has been prepared within frameworks of the planar technology on the n-type 6H-SiC (0001) wafer doped with nitrogen at the concentration of $2 \cdot 10^{18}$ cm$^{-3}$. Firstly, the pyrolysis of silane was used to obtain the oxide overlayer on the 6H-SiC (0001) surface. Then, the photolithography processes and subsequent etching were applied to make the Hall geometry windows with sizes of 4.72 mm x 0.20 mm (see Fig. 1). Finally, the short-time diffusion of boron has been performed from the gas phase under controlled injection of the silicon vacancies at the temperature of T = 900$^0$C [2].

The quantum conductance, tunneling spectroscopy and SIMS measurements showed that the p+-diffusion profile obtained represents the ultra-narrow p-type quantum well (QW), 2 nm, confined by the δ-barriers heavily doped with boron up to N(B) > $10^{21}$ cm$^{-3}$, 3 nm (see Fig. 1). Besides, the δ-barriers heavily doped with boron appeared to be the effective broadband source of the GHz and THz emission under the voltage applied both along the quantum well and to the p$^+$-n junction, because contain predominantly high concentration of the negative-U dipole boron centers [3]. The frequency selection that is controlled by measuring the Shapiro steps became it possible, because the special microcavities have been incorporated into the QW plane, with the sizes corresponding to the frequencies noticed above, L=λ/2n; here n is the refractive index, L is the microcavity size and λ is the wavelength of the GHz- and THz-radiation [3]. Here the length of the δ-barrier area, L=4.72 mm, was picked up to satisfy to the selection of the frequency value of 9.3 GHz.

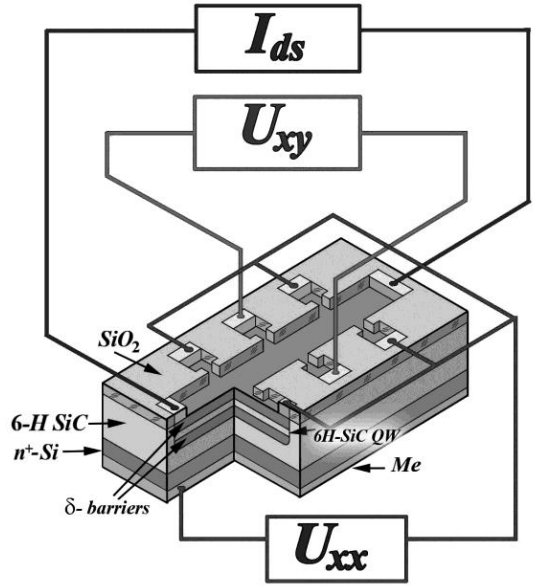

**FIGURE 1.** Device schematic, showing the perspective view of the 6H-SiC sandwich structure performed within the frameworks of the Hall geometry. The 6H-SiC sandwich structure represents the p-type 6H-SiC quantum well confined by the δ-barriers heavily doped with boron on the n-type 6H-SiC surface. The changes of the longitudinal, $U_{xx}$, and the transverse, $U_{xy}$, voltage are measured by varying the value of the external magnetic field when the drain-source current, $I_{ds}$, is stabilized at the extremely low value.

Thus, the GHz emission from the δ-barriers appeared to be a basis of the EDESR technique by measuring the only dependences of the longitudinal, $U_{xx}$, and transverse, $U_{xy}$, voltages on the magnetic field value without the external hf source and recorder as well as the external cavity [3]. Finally, the ESR measurements were also performed with the Bruker ELEXSYS E580 spectrometer at 9.7 GHz in the temperature interval from 5 K to 40 K.

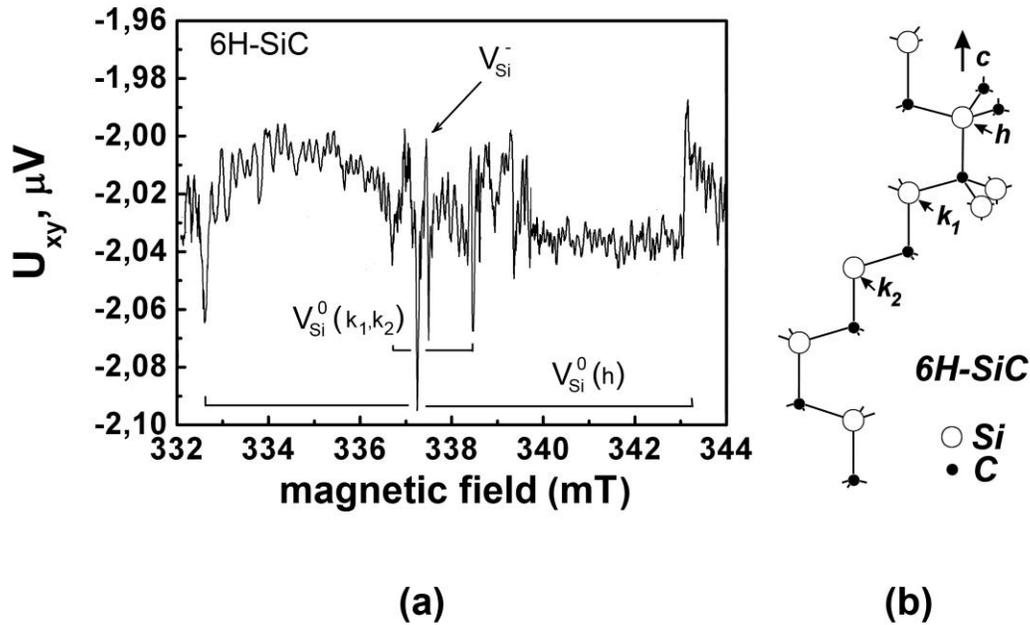

**FIGURE 2.** EDESR spectra of silicon vacancy centers in the p-type 6H-SiC-QW confined by the δ-barriers heavily doped with boron on the n-type 6H-SiC (0001) surface, which is observed by measuring a magnetoresistance without the external cavity as well as the external hf source and recorder. B∥c, T=77 K, ν=9.3GHz, $I_{ds}$=10 nA

# RESULTS AND DISCUSSION

The EDESR spectra of the silicon vacancy centers that have been formed in the process of the preparation of the 6H-SiC-QW confined by the δ-barriers heavily doped with boron are shown in Fig. 2a. These spectra have been already noticed to be observed by measuring the only magnetoresistance without the external cavity as well as the hf source and recorder. T=77 K. B ∥ c. ν =9.3 GHz. $I_{ds}$ = 1 nA. The results are in a good agreement with the data obtained by the X-band ESR method in the studies of the silicon vacancy centers created under e-irradiation in the 6H-SiC bulk sample [4]. In particular, the central line in Fig. 2a marked by $V_{Si}^-$ is attributed to the negatively charged silicon vacancy with S = 3/2, whereas the lines indicated by $V_{Si}^0(h)$ and $V_{Si}^0(k_1,k_2)$ are related to the neutral charge silicon vacancy but in different positions in crystalline lattice. Symbols "h" and "$k_1$, $k_2$" depict the hexagonal and the quasicubic positions in the 6H-SiC crystal that has a hexagonal symmetry with symmetry axis c (see Fig. 2b). It should be noted that the different phase of the lines of the triplet $V_{Si}^0(h)$ is evidence of non-equilibrium spin polarization that seems to be caused by the hf frequency pumping from the δ-barriers heavily doped with boron.

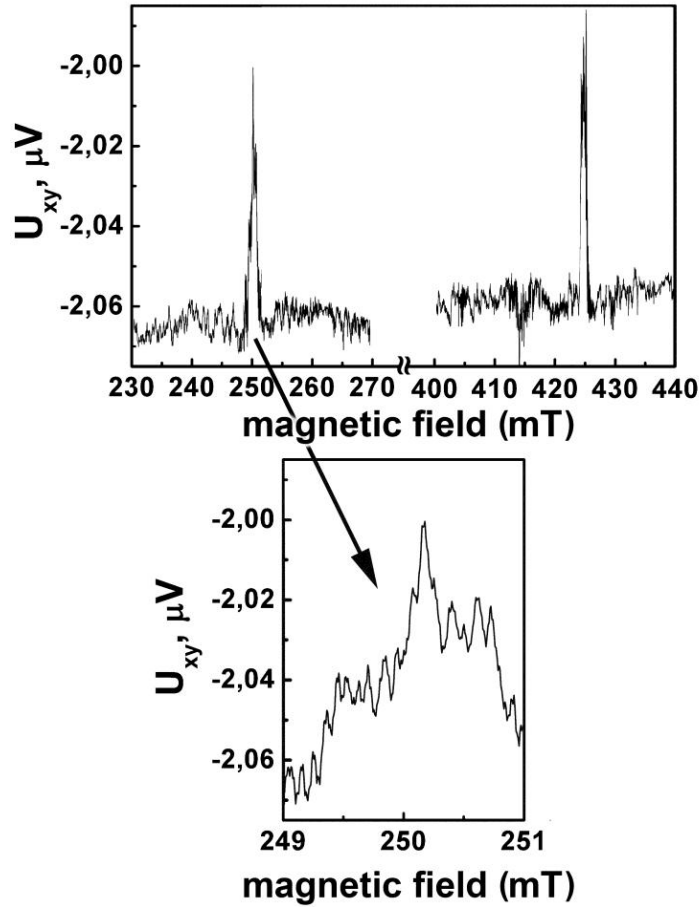

**FIGURE 3.** EDESR spectrum of N-$V_{Si}$ defect (S=1) observed in the p-type 6H-SiC-QW confined by the δ-barriers heavily doped with boron on the n-type 6H-SiC (0001) surface, which is observed by measuring a magnetoresistance without the external cavity as well as the external hf source and recorder. B∥c, T=77 K. ν=9.3GHz, $I_{ds}$=10 nA. Fine structure resolved in the EDESR spectrum appears to be evidence of the nitrogen presence (I=1). Isolated nitrogen donor centers inside the p-type 6H-SiC-QW are absent as a result of interaction with silicon vacancies.

The EDESR spectrum of the triplet N-$V_{Si}$ defect (S=1) consists of two strong lines located at 250 mT and 450 mT (Fig. 3). The high resolution of the EDESR method allowed the registration of the N-$V_{Si}$ defect fine structure as a result of which the lines of spectra are divided into three lines due to nitrogen nuclear spin I=1, with the value of the hf splitting equal to 0.5 mT. The absence of the ESR lines related to the nitrogen donor center is of importance to be noticed, which seems to result from the capture of silicon vacancies practically on all of them in the area of the quantum well confined by the δ-barriers heavily doped with boron. Besides, the sensitivity volume of the EDESR method is about several nanometers in the depth of the sample multiplied on Hall bar geometry area of the device structure that hinders with the identification of defects in the volume of the n-type 6H SiC (0001) wafer. However the EDESR method based on the registration of the negative or positive magnetoresistance in the weak localization regime under resonance conditions is highly sensitive inside the QW area [3]. In particular, the EDESR spectrum of the N-$V_{Si}$ defect shown in Fig. 3 reflects the contribution from 5 $10^4$ centers taking account of the section area of the edge channels in the device structure, 2 nm x 2nm, while the ordinary ESR method allows signal detection minimum from $10^{10}$ spins.

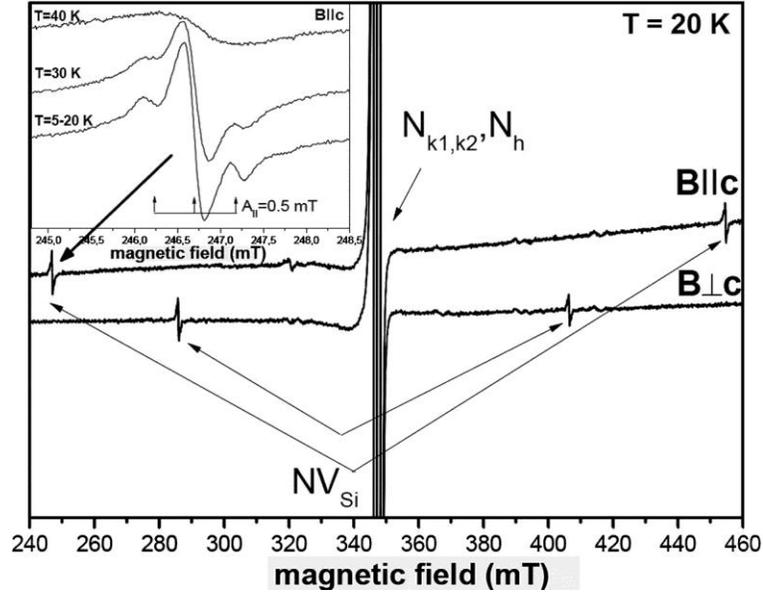

**FIGURE 4.** The ESR spectrum of the triplet center observed in the p-type 6H-SiC-QW confined by the δ-barriers heavily doped with boron on the n-type 6H-SiC (0001) surface. The hf structure resolved in the ESR spectrum of a triplet center was shown in insertion. T=20K.

This value corresponds to the maximum number of the N-$V_{Si}$ defects which can be created by the capture of silicon vacancies on nitrogen donors in the process of the preparation of the device if to take into account the sizes of the doping area. This important condition allowed the ESR detection of the N-$V_{Si}$ defect in the same device with the highly sensitive Bruker ELEXSYS E580 spectrometer (Fig. 4).

In addition to the dominant spectrum of the nitrogen (N) donors substituting quasi-cubic ($N_{k1,k2}$) and hexagonal ($N_h$) sites originated from the n-type 6H-SiC (0001) bulk area, a fine structure (fs) doublet lines of small intensity with fs splitting of $\Delta B_{\parallel}$=237.6 mT is observed in the temperature interval from 5 K to 40 K and related to the triplet center with $S$=1. The fact that the fs lines corresponding to the electron transitions ($M_S$=1→$M_S$=0; $M_S$=−1→$M_S$=0) were detected in darkness confirms a triplet ground state of the defect center. When the amplitude modulation of the magnetic field was taken smaller than the width of the triplet lines, as can be seen from Fig. 4, three hyperfine (hf) lines can be well resolved in the ESR spectrum of the defect center with value of the hf splitting of about 0.5 mT. The presence of the three hf lines indicates its origin from the hf interaction with one $^{14}$N nuclei ($S$=1/2 and $I$=1). Thus, triplet center seems to consist of the nitrogen atom and silicon vacancy introduced into the device in the process of its preparation. Since there are three nonequivalent positions for the impurities and defects in the 6H-SiC lattice, it is expected that this property is also attributed to the observed defect. Therefore different intensity of the central hf triplet line and sidelines in Fig. 4 can be explained by the contribution of the ESR spectrum due to the defect substituting hexagonal site in the central line, which has smaller unresolved hf structure. From the ratio between the central and sidelines of the nitrogen triplet, it was concluded that the triplet with the hf splitting of the 0.5 mT value corresponds to the defect substituting one quasi-cubic position. The angular dependence of the ESR spectrum shown in Fig. 4 that was measured by the rotation of the magnetic field in the plane consisting of the **c** axis appears to demonstrate the angular variation characteristic for a spin $S$=1.

For comparison, the parameters of the N-$V_{Si}$ triplet centers containing in their structure nitrogen atoms which were observed previously in electron (e) -irradiated (dose $10^{12}$e/cm$^2$-$10^{15}$e/cm$^2$) and annealed at T=900$^0$C synthetic diamond and in neutron (n) -irradiated (dose of $10^{21}$ cm$^{-2}$) n-type 6H-SiC after high temperature annealing at T=2000$^0$C were listed in Table 1.

**TABLE 1.** Parameters of the spin Hamiltonian for triplet centers containing N atoms observed in the 6H-SiC nanostructures, e-irradiated synthetic diamond and n-irradiated 6H-SiC of the n-type.

| Center | S | T [K] | g | A [mT] | D, 10$^{-4}$ [cm$^{-1}$] | Ref. |
|---|---|---|---|---|---|---|
| N-V Diamond e-irradiated | 1 |  | $g_{iso}$ = 2.0028 | $A_\parallel$ = 0.082 $A_\perp$ = 0.075 | 960.7 | [5] |
| N$_C$-V$_{Si}$ 6H-SiC n-irradiated | 1 | 3.5 | $g_{iso}$ = 2.003 | $A_\parallel$ = 0.55 | 860 | [6] |
|  |  | 70 |  | $A_\parallel$ = 0.75 $A_\perp$ = 0.35 | 885 |  |
| N$_C$-V$_{Si}$ 6H-SiC nanostructure | 1 | 5-40 | $g_\parallel$ = 1.9700(3) $g_\perp$ = 1.9961(3) | $A_\parallel$ = 0.55 $A_\perp$ = 0.45 | 1140 | This work |

As can be seen from Table 1, the observed triplet centers have close values of the fs splitting constants and were attributed to the defect center formed between nitrogen atom and vacancy. The triplet center observed in the 6H-SiC nanostructure has larger fs constant than that observed in irradiated synthetic diamond and 6H-SiC. At the same time the hf splitting due to the interaction with $^{14}$N nucleus has the same value as for the triplet center observed in the n-irradiated 6H-SiC at 5 K. Taking into account that silicon vacancy (V$_{Si}$) is more mobile at 900$^0$C than the V$_C$ and appears to dominate as a carrier in the process of low temperature impurity diffusion into the n-type 6H-SiC bulk samples, the observed triplet center was also attributed to the N-V$_{Si}$ –related center.

Finally, the striking result obtained in this work is that in distinction from the known N-V$_{Si}$ defects observed in the e-irradiated diamond and in the heavy n-irradiated and high temperature annealed n-type 6H-SiC, the N-V$_{Si}$ center characterized by the larger value of the zero-field splitting constant D and anisotropic g-factor was found in the non-irradiated 6H-SiC nanostructure.

## ACKNOWLEDGMENTS

The work was supported by the programme of fundamental studies of the Presidium of the Russian Academy of Sciences "Quantum Physics of Condensed Matter" (grant 9.12); programme of the Swiss National Science Foundation (grant IZ73Z0_127945/1); the Federal Targeted Programme on Research and Development in Priority Areas for the Russian Science and Technology Complex in 2007–2012 (contract no. 02.514.11.4074), the SEVENTH FRAMEWORK PROGRAMME Marie Curie Actions PIRSES-GA-2009-246784 project SPINMET and . SAFMAT project CZ.2.16/3.1.00/22132.

## REFERENCES

1. D. Riedel, F. Fuchs, H. Kraus, S. Väth, A. Sperlich, V. Dyakonov, A. A. Soltamova, P. G. Baranov, V. A. Ilyin, and G. V. Astakhov, *Phys. Rev. Letters* **109**, 226402 – 06 (2012).
2. N. T. Bagraev, A. A. Gippius, L. E. Klyachkin, A. M. Malyarenko, *Mater. Sci. Forum* **258-263**, 1833-1838 (1997).
3. N. T. Bagraev, V. A. Mashkov, E. Yu. Danilovsky, W. Gehlhoff, D. S. Gets, L. E. Klyachkin, A. A. Kudryavtsev, R. V. Kuzmin, A. M. Malyarenko, V. V. Romanov, *Appl. Magn. Resonance* **39**, 113–135 (2010).
4. P. G. Baranov, A. P. Bundakova, A. A. Soltamova, S. B. Orlinskii, I. V. Borovykh, R. Zondervan, R. Verberk, and J. Schmidt, *Phys. Rev. B* **83**, 125203-12 (2011).
5. X.-F. He, N. B. Manson, P. T. H. Fisk, *Phys. Rev B* **47,** 8816-8822 (2000).
6. M. V. Muzafarova, I. V. Ilyin, E. N. Mokhov, V. I. Sankin, P. G. Baranov, *Mater. Sci. Forum* **527-529,** 555-529 (2006).